\documentclass[twocolumn,secnumarabic,amssymb, nobibnotes, aps, prb]{revtex4}
\usepackage{graphicx}
\usepackage{amsmath}
\usepackage{bm}
\begin{document}

\title{ Lattice structures of Larkin-Ovchinnikov-Fulde - Ferrell (LOFF) state  }%

\author{Longhua Jiang and Jinwu Ye }%
\affiliation{  Physics Department, The Pennsylvania State University, University Park, PA, 16802 }
\date{\today}%

\begin{abstract}
 Starting from the Ginzburg-Landau free energy describing
 the normal state to Larkin-Ovchinnikov-Fulde-Ferrell (LOFF) state
 transition, we evaluate the free energy of seven most common
 lattice structures such as stripe, square, triangular,
   Simple Cubic (SC), Face centered Cubic (FCC),
   Body centered Cubic (BCC) and Quasi-crystal (QC). We find that  the stripe phase
  which is the original LO state, is the most stable phase.
  This result maybe relevant to the detection of LOFF state in some heavy fermion compounds and
  the pairing lattice structure of fermions with unequal populations in the BCS side of Feshbach resonance in
  ultra-cold atoms.
\end{abstract}
\maketitle

\section{ Introduction }

   It is well known that at sufficiently low temperature, an electron with spin up is
   paired with its partner with spin down across the Fermi surface to
   form a Cooper pair with total momentum zero and becomes superconductor and exhibits superfluid property.
   This phenomenon is well described by Bardeen-Copper-Schrieffer ( BCS ) theory.
   The most favorable condition for paring is when spin up and spin down electrons have the same density.
   Now imagine one apply a magnetic field to split the spin up and spin down electrons by
   Zeeman effect and look at the response of a superconductor to the
   Zeeman splitting.
   For $ s $-wave superconductor, if the Zeeman splitting
   $ \delta \mu = \mu_{\uparrow}-\mu_{\downarrow} $ is very small compared to the gap,
   then the  superconducting state is stable, if it is much larger than the
   gap, the superconducting state will turn into a normal state.
   When $ \delta \mu $ is comparable to the energy gap $ \Delta_{0} $ at zero magnetic
   field, it may becomes non-trivial. It was argued by Fulde and Ferrell \cite{ff},
   Larkin and Ovchinnikov \cite{lo} about 40 years ago that an in-homogeneous
   superconductor with pairing order parameter oscillating in space
   may be the ground state at a narrow window of Zeeman splitting
   $ \delta \mu_{1} \sim \Delta_{0}/\sqrt{2} < \delta \mu <  \delta \mu_{2} \sim 0.754
    \Delta_{0} $ \cite{ms,loff} ( Fig.1 ). This in-homogeneous state is called LOFF state where
    the Cooper pairs carry a finite momentum. In FF state, $ \Delta(x) =
    \Delta_{0} e^{i \vec{q} \cdot \vec{x}} $ where $ q \sim k_{F \uparrow} -k_{ F\downarrow} $, the
    Cooper pairs carry finite superfluid momentum ,
    while in the LO state, $ \Delta(x) =  \Delta_{0} \cos \vec{q} \cdot \vec{x}
    $, the Cooper pairs carry two opposite momenta.
    The LOFF state breaks both $ U(1) $ gauge symmetry  and
   translational order. Unfortunately, so far, the
   LOFF state has never been observed in conventional
   superconductors, because in these systems,  the Zeeman effect is overwhelmed by orbital effects.
   However, this LOFF state has attracted renewed interests in the context of
   organic, heavy fermion and high $ T_{c} $ cuprates \cite{exp,heavy}, because
   these new classes of superconductors may provide favorable conditions
   to realize the LOFF state.
   Recently, experiments \cite{martin} on penetration depth measurement on $CoCeIn_{5}$ shows
   that at a temperature below 250 mK, for magnetic field applied  parallel to the $ab$ plane, two phase
   transitions were detected, one of which maybe identified as a phase transition from LOFF state to normal state
   transition. Also the measurement of thermal conductivity\cite{capan} on $CoCeIn_{5}$
   shows anisotropy in real space, which could be interpreted as domain wall formation, namely, a stripe phase
   but possibly with higher harmonics. LOFF states also played important roles in high density quark matter,
   astrophysics \cite{loff} and superconductor-ferromagnet heterostructures \cite{jun}.
   With the development of trapped cold atoms system, it was proposed that due to absence of
   orbital effects, ultracold neutral fermion gases with unequal populations may realize the LOFF state
   in a tiny window on the BCS side of Feshbach resonance \cite{fesh}.
   Recently, it was argued in \cite{yip} that the LO state, in fact,
   may be stable in an appreciate regime in the BCS side of the
   Feshbach resonance.

\begin{figure}
\includegraphics[width=2.5in]{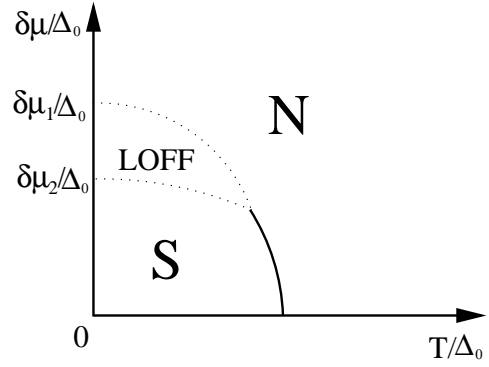}

\caption{The phase diagram of LOFF state. $ \delta \mu $ is the
Zeeman splitting, $ T $ is the temperature, $ \Delta_{0} $ is the
energy gap at the balanced case $ \delta \mu=0 $. }
\end{figure}

   Before we discuss the phase diagram Fig.1, we reviewed the basic facts of classical Lifschitz point
   which is closely related to normal state to LOFF state phase transition.
   This connection is not that new, but has not been stressed in any literature.
   The free energy near a classical $ ( d, d_{\perp} ) $ Lifshitz
   point is \cite{tom}:
\begin{eqnarray}
   H&=& \frac{1}{2} \int d^{d} x [ t m^{2} + K_{\parallel} ( \nabla m )^{2} +
   K_{\perp} ( \nabla m )^{2}  \nonumber \\
   &+& L ( \nabla^{2} m )^{2} ]+ u  \int d^{d} x m^{4} + \cdots
\label{first}
\end{eqnarray}
    where $ K_{\parallel} > 0 $ and  $ m( x ) $ is a $ n \geq 2 $ component order parameter,
    the dimension $ d $ is divided into $ d_{\perp} $ perpendicular dimension and
    $ d_{\parallel} $ parallel dimension. Its phase diagram \cite{tom} is shown in Fig.2.
\begin{figure}
\includegraphics[width=3.5in]{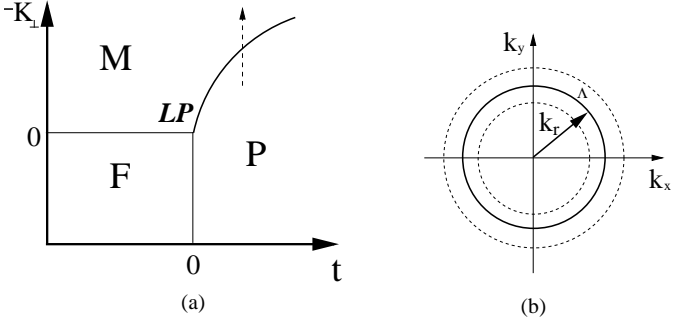}
\caption{\footnotesize (a) Phase diagram of Classical Lifshitz point
(CLP). P is the Paramagnetic phase, F is the ferromagnetic phase, M
is the modulated phase. The LP point is at  $ (t, K_{\perp} )=(0,0)
$. The dashed line is the P-M transition we are studying. (b)
Momentum shell of width $ \Lambda $ around 2d roton surface. }
\end{figure}

   Let me review the phase transition from $ P $ to $ M $ transition along the dashed line shown in
   Fig.2. In the P phase along the path close to the P-M transition boundary,
   $ t >0, K_{\parallel}>0, K_{\perp} < 0 $, for simplicity, we can set $
   k_{\parallel}=0 $,
   the propagator $ D(k_{\parallel}=0, k_{\perp}) $ can be
   written as $ D(k_{\perp})= t+K_{\perp}  k^{2}_{\perp} + L k^{4}_{\perp} = \Delta
   + L ( k^{2}_{\perp}- k^{2}_{r} )^{2} $
   where $ \Delta= t- \frac{ K^{2}_{\perp} }{4 L }, k^{2}_{r}=
  \frac{|K_{\perp}|}{2L} $.  It is easy to see
   the minima is located at the  " roton " surface $ k^{2}_{r} $ ( Fig. 2b), in
   sharp contrast to $ K_{\perp} > 0 $
   case where the minimum is at $  k_{\perp}=0 $. This class of problems with minima located at $ k_{r} > 0 $
   was first investigated in \cite{bs} and has wide applications in
   the context of liquid crystals \cite{tom}. When $ \Delta
    > 0 $, the system is in the paramagnetic ( P ) phase with $ < m > =0 $, while when $
   \Delta < 0 $, it is in a modulated ( M ) phase with the mean
   field structure $ < m( x )> = \sum^{P}_{i=1} \Delta_{i}
   e^{i \vec{p}_{i} \cdot \vec{x} }, \ q_{i}= k_{r} $. The $ P-M $
   transition happens at $ \Delta=0 $, namely, $ t= \frac{ K^{2}_{\perp} }{4 L
   } $ as shown in Fig. 2. The $ M $ phase breaks
   both the internal $ O(n) $ rotational symmetry and
   the translational symmetry, therefore supports two kinds of Goldstone modes:
   phase mode due to the $ O(n) $ symmetry breaking and the lattice phonon mode
   due to the translational symmetry breaking. At the mean field theory, the P-M
   transition is 2nd order.  Under fluctuations,
   For $ d_{\perp} =1 $, the roton surface  in Fig.2b, in fact, turns into two isolated
   points, the transition which describes nematic-Smectic A transition in liquid crystal
   remains 2nd order. However, for $ d_{\perp} \geq 2 $, the transition becomes a fluctuation driven
   1st order transition as shown by Renormalization group  analysis in \cite{qgl}.
   Indeed, to some extent, the LOFF
   phase diagram Fig.2 looks similar to Fig. 1 if we identify Zeeman splitting
   $\delta \mu $ as the pressure $ -K_{\perp} $, normal phase as the
   paramagnetic phase, the superconducting phase as the
   ferromagnetic phase and the LOFF state as the modulated phase.

   Of course, the original pairing problem of fermions with unequal
   populations are a fermionic problem. However, just like usual
   normal state to BCS superconductor transition, one can integrate
   out fermions at any finite temperature and lead to
   the following Ginsburg-Landau free energy describing the normal state to the LOFF state
   transition \cite{Hou1,Hou2,loff,kun}:
\begin{eqnarray}
& f &\propto|(-\nabla^2-q^2_{0} )\psi|^2+a|\psi|^2+b|\psi|^4+c|\psi|^2|\nabla\psi|^2 \nonumber \\
&+&d[(\psi^*)^2(\nabla\psi)^2+\psi^2(\nabla\psi^*)^2]+e|\psi|^6,
\label{f}
\end{eqnarray}
    where  $ q_{0} \sim k_{F \uparrow}-k_{F \downarrow} $.

    Indeed, this action is very similar to the Lifshitz action
    Eqn.\ref{first} with $ K_{\perp} < 0 $, so similar procedures
    following Eqn.\ref{first} can be used.
    Substituting $ \psi=\sum_{G}\psi_{G}e^{iGx} $ where $ G $ are
    the shortest reciprocal lattice vectors into the above
    equation and combining terms lead to the GL free energy
    in momentum space:
\begin{eqnarray}
 f &= & \sum_{G}\frac{1}{2}r_{G}|\psi_{G}|^{2}+u\sum_{G}\psi_{G_{1}}\psi_{G_{2}}\psi_{G_{3}}\psi_{G_{4}}\delta_{G_{1}+G_{2}+G_{3}+G_{4}}\nonumber \\
&+&v\sum_{G}\psi_{G_{1}}\psi_{G_{2}}\psi_{G_{3}}\psi_{G_{4}}\psi_{G_{5}}\psi_{G_{6}}\delta_{G_{1}+
G_{2}+G_{3}+G_{4}+G_{5}+G_{6}}
\label{mom}
\end{eqnarray}
   where $ r= T-T_{c} $ and $ u,v $ are functions of the coefficients $ b,c,d,e $ in Eqn.\ref{f} and  $\vec{G} $.

   If $ r > 0 $,  the system is in the normal state with $ < \psi(\vec{G}) > =0 $,
   while when $ r < 0 $, it is in a modulated ( M ) phase with the mean
   field structure $ < \psi ( x )> = \sum^{P}_{i=1} \Delta_{i}
   e^{i \vec{q}_{i} \cdot \vec{x} }, \ q_{i}= q_{0} $. This $ M $
   phase is the LOFF state.
   The LOFF state breaks both $ U(1) $
   symmetry and the translational symmetry, therefore it supports two kinds of Goldstone
   modes. (1) the Goldstone mode due to the $ U(1) $ symmetry
   breaking, but it was "eaten" by the gauge field due to Higgs
   mechanism in electron  pairing case in condensed matter system, but will stay in
   the neutral atom pairing case in ultra cold atom atomic experiments  (2) the lattice phonon modes due to the translational symmetry
   breaking, they will survive the gauge field fluctuations. In this
   paper, we approach the LOFF state from the normal state and try
   to determine what is the lowest lattice structure of the LOFF
   state. $ P=1 $ corresponds to the FF state, $ P=2 $ corresponds
   to the LO state. It is known that the FF state, being carry
   finite superfluid momentum, is always unstable. The LO state has
   nodes where the excess fermions reside. However, it is still not know the LO state is the most
   favorable lattice structure. In this paper, we will study what is
   the lowest lattice structure by
   considering seven most common lattice structures namely the stripe, square, triangular,
   Simple Cubic (SC), Face centered Cubic (FCC),
   Body centered Cubic (BCC) and Quasi-crystal (QC) listed in Table I. The stripe case
   corresponds to the original LO state.

   The rest of the paper is organized as follows.
   In  section II, we compute the coefficients of the free energy of the LOEF states with different lattice structures.
   In section III,  by comparing the free energy and the transition
   temperature of all the seven lattice structures of LOFF state,
   we find the lowest energy lattice structure remains the LO state.
   In the appendix A, we discuss in detail how to get the geometrical factors in the
   fourth and sixth order terms which are used in evaluating the
   free energy of the seven lattices. As a byproduct, we corrected some over-counting mistakes
   in describing liquid to solid transition in the textbook in \cite{tom}. In appendix B, we revisit the
   solid to liquid transition by considering both cubic and quartic
   term and show that the BCC lattice  remains the favorable
   lattice in the presence of cubic term in a certain region.

\section{ Effective free energies of the LOFF state with different lattice structures }

  We only look at the subset $ L_{G} $ spanned by all the shortest reciprocal lattice vectors $ G=q_{0}
  $. In the ground state, $ \psi_{G} $ has to be real up to a global
  phase. From the point group symmetry of the lattices, $ \psi_{G} $
  is a constant when $ G $ belongs to $ L_{G} $. Following
  \cite{tom}, we have scaled $ n_{G} \rightarrow  n_{G} m^{-1/2} $
  so the quadratic term is the same for all the lattices. Then Eqn.\ref{mom}
  is simplified to the effective free energy in different lattices:
\begin{equation}
f=\frac{1}{2}r\psi_{G}^{2}+u_{\alpha}\psi_{G}^{4}+v_{\alpha}\psi_{G}^{6}
\label{eff}
\end{equation}
     where $ \alpha $ stands for different lattices. In the
     following, we will calculate the fourth order term $ u_{\alpha}$ and the sixth order term $ v_{\alpha} $ for
     different lattices respectively.

{\sl 1. The fourth order term $ u_{\alpha} $. } For stripe phase,
square lattice ,triangular lattice, SC and FCC, as shown in the
appendix A, there are only  contributions from paired vectors to the
quartic term $u^{p}_{\alpha}=3(1-\frac{1}{m})$ where $ m $ is number
of the vectors in the set $L_{G}$. Therefore
$u_{\Vert}=\frac{3}{2}u, u_{\Box}=\frac{9}{4}u,
  u_{\triangle}=\frac{5}{2}u, u_{sc}=\frac{5}{2}u,
  u_{fcc}=\frac{21}{8}u $. The set $ L_{G} $ for
different lattices are shown in Fig.3 for one and two dimensional
lattices and Fig.4 for three dimensional lattices.

\begin{figure}
\includegraphics[width=3.5in]{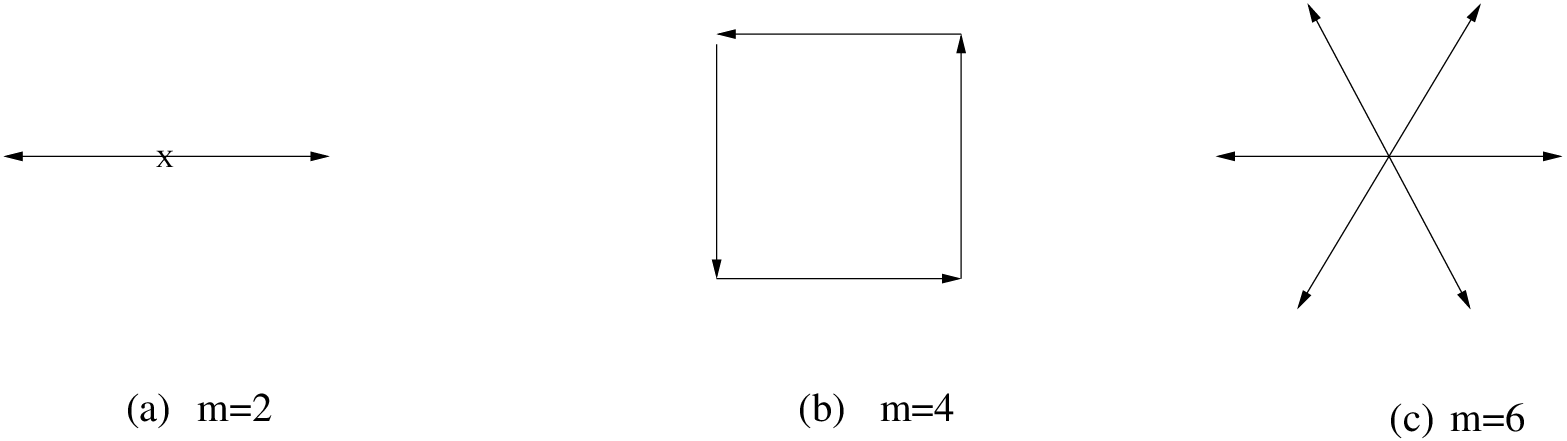}
 \caption{\footnotesize The set of shortest reciprocal lattice vectors  $ L_{G} $ for one and two dimensional
 lattices (a)
  Stripe lattice (b)  Square lattice (c)  Triangular lattice}
\end{figure}

\begin{figure}
  \includegraphics[width=3.5in]{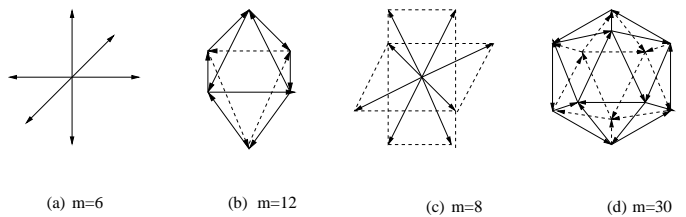}
  \caption{\footnotesize The set of shortest reciprocal lattice vectors  $ L_{G} $ for three dimensional
  lattices (a) Simple Cubic (b) BCC lattice (c)  FCC lattice (d)  Quasicrystal }
\end{figure}

  For a BCC lattices, there is an additional
  vertex contribution $ u_{v}= u $ coming from the 4 vectors from any of the six vertices.
  So in all, $u_{bcc}=u_{p}+u_{v}= \frac{15}{4}u $.

  For a quasi-crystal, we have an additional contribution from the non-planar diamonds
  \cite{tom} $ u_{npd}=\frac{4}{5}u $, so in all,
  $ u_{qc}= u_{p}+ u_{npd}= \frac{37}{10}u $.

{\sl 2. The sixth order term $ u_{\alpha} $ }
   For the stripe
  phase, square lattice, SC and FCC, there are only contributions from paired
  vectors $v^{p}_{\alpha}=5({3m^{2}-9m+8})/ m^{2}$. So we get
 $v_{\Vert}=2\frac{1}{2}v, v_{\Box}=6\frac{1}{4}v,
 v_{sc}=\frac{155}{18}v, v_{fcc}=10v $.

 For the triangular lattice, there is an additional contribution
 $ v_{tri}=\frac{5}{6}v$ coming from the
 closed triangles diagram ( Fig.5c ). So we get $
 v_{\triangle}=v_{p}+v_{tri} = 9 \frac{4}{9} v$.

 For the BCC,  in additional to the paired vector contributions
 $ v^{p}= \frac{415}{36}v$, there are also contributions coming from the
 three configurations listed in Fig.5 which is $  \frac{155}{12} v $.
 In all, $ v_{bcc}=220/9 v $.

 For Quasicrystal, in additional to the paired vector contributions
 $ v^{p}=\frac{1219}{90}v $, there are also contributions coming from the
 four configurations listed in Fig.6 which is $  \frac{211}{15}v$.
 In all, $ v_{qc}=497/18v $

\begin{figure}  \includegraphics[width=3.5in]{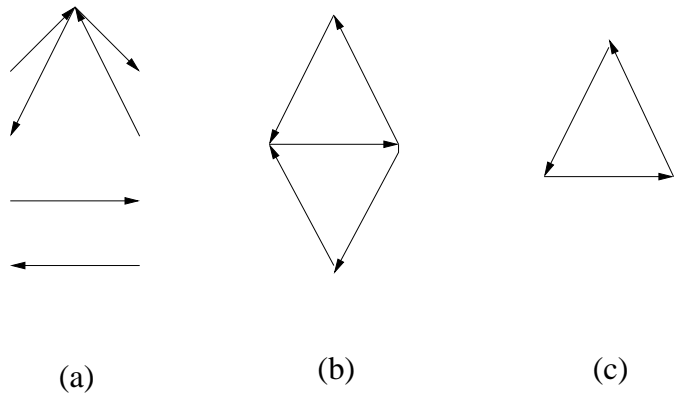}
  \caption{\footnotesize non-paired contributions to sixth order term in BCC lattice
  (a) a pair of opposite vectors plus four vectors coming out of one vortex,10v,
  (b) a non-planar triangle diagram with the common edge chosen twice, $\frac{5}{2}v$
  (c) a triangle diagram, each vector in the triangle was chosen twice, $\frac{5}{12}v $;
      for the triangle lattice in Fig.3c, this term is $ \frac{5}{6}v $ }
\end{figure}
\begin{figure}
 \includegraphics[width=3.5in]{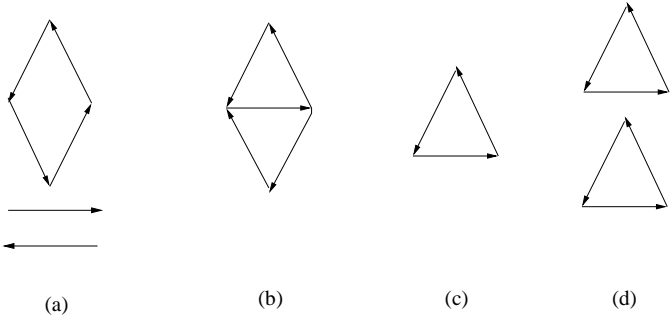}
 \caption{\footnotesize non-paired contributions to sixth order term in Quasicrystal lattice
  (a) a pair of opposite vectors plus a non-planar diamond structure, $\frac{52}{5}v$.
  (b) a non-planar triangle diagram with the common edge chosen twice, $\frac{2}{5}v$.
  (c) a triangle diagram, each vector in the triangle was chosen twice, $\frac{1}{15}v$.
  (d) two triangles with no common edges, $\frac{16}{5}v$ }
\end{figure}

   The $ u_{\alpha} $ and $ v_{\alpha} $ for the seven lattices are
   listed in the following table. \\
 \begin{table}[htbp]

   \begin{tabular}{|c|c|c|c|c|c|c|c}\hline
    lattices & stripe &square&triangular&SC&BCC&FCC&QC\\ \hline
  $u_{\alpha}$   & $\frac{3}{2}u$ &$\frac{9}{4}u$ &$\frac{5}{2}u$&  $\frac{5}{2}u$ &$\frac{15}{4}u $&$\frac{21}{8}u$&$\frac{37}{10}u$   \\ \hline
  $v_{\alpha}$   &$\frac{5}{2}v$ &$\frac{25}{4}v$ &$\frac{85}{9}v$&  $\frac{155}{18}v$ &$\frac{220}{9}v$&$10v$&$\frac{497}{18}v$\\ \hline
   \end{tabular}
 \caption{ $ u $ and  $ v $ for the seven lattices}
 \end{table}
\\

\section{ Optimal lattice structure of the LOFF state }

In the original GL action Eqn.\ref{mom}, $ u $ can be negative and
positive.  In case $ v $ is also negative, then an eighth order is
needed. In this paper, we assume $ v $ is always positive to keep
the system stable. In the following, we discuss $ u < 0 $ and $ u >
0 $ cases respectively.\\
{\sl 1. $ u $ is positive. }
 It is easy to see that $u_{\Vert}<u_{\Box}<u_{sc}=u_{\triangle}<u_{fcc}<u_{bcc}$ and $v_{\Vert}<v_{\Box}<v_{sc}<v_{\triangle}<v_{fcc}<v_{bcc}$  so for any given $ \psi $:  $
f_{\Vert}(\psi)<f_{\Box}(\psi)<f_{sc}(\psi)<f_{\triangle}(\psi)<f_{fcc}(\psi)<f_{bcc}(\psi)
$. Then $
f_{\Vert}(\psi_{\Vert})<f_{\Box}(\psi_{\Box})<f_{sc}(\psi_{sc})<f_{\triangle}(\psi_{\triangle})<
f_{fcc}(\psi_{fcc})<f_{bcc}(\psi_{bcc}) $.
 However, more work is
needed to compare  Quasicrytal with BCC. Minimization of
Eqn.\ref{eff} leads to the order parameter and the free energy:
\begin{eqnarray}
\psi_{\alpha}^{2} & = &
 \frac{-2u_{\alpha}+\sqrt{4{u^{2}_{\alpha}-6v_{\alpha}r}}}{6v_{\alpha}}
    \nonumber \\
f & = &
\frac{6rv_{\alpha}-4u^{2}_{\alpha}}{18v_{\alpha}}\psi_{\alpha}^{2}-\frac{u_{\alpha}r}{18v_{\alpha}}
\label{free}
\end{eqnarray}

     Defining $ r=x \frac{u^{2}}{v}$ where $ x $ is dimensionless and plugging it into Eqn.\ref{free},
     we get $ f_{\alpha}=\frac{u^{3}}{v^{2}} g_{\alpha}(x) $ where $ g_{\alpha} $ are dimensionless
     functions and $\alpha$ stands for Quasicrytal and BCC. Comparing these two functions,
     we find that there is a shift of order between these lattices as shown in
     Fig.7.

     \begin{figure}
 \includegraphics[width=3in]{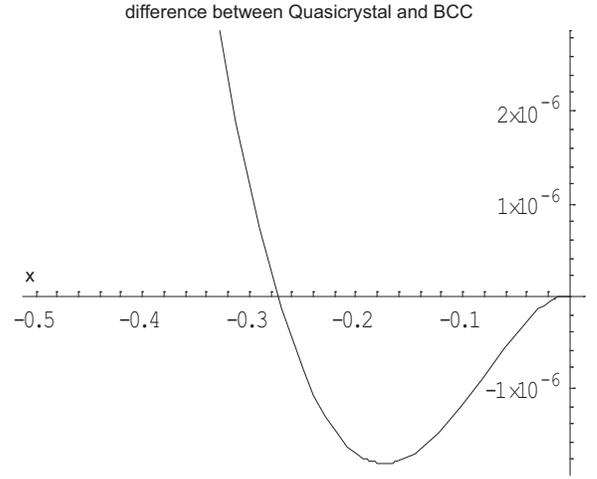}
 \caption{ $ u $ is positive. Difference between $ g_{qc} $ and $g_{bcc}$. }
  \end{figure}

     When $-0.274\frac{u^{2}}{v}<r<0$, $g_{qc}<g_{bcc}$ thus $f_{qc}<f_{bcc}$.
     However when $ r<-0.274\frac{u^{2}}{v} $, $ g_{qc}>g_{bcc}$ thus $ f_{qc} > f_{bcc}$.
     In any case, the stripe phase is the lowest free energy lattice.\\
{\sl 2. $ u $ is negative. } Eqn. \ref{free} still hold for $ u<0 $.
 We can use the same method used when u is positive.  Defining $
r=x \frac{u^{2}}{v}$ and plugging it into Eqn.\ref{free}, we still
have the following expression $
f_{\alpha}=\frac{u^{3}}{v^{2}}g_{\alpha}(x)$. For seven different
lattices, we get the same coefficient $\frac{u^{3}}{v^{2}}$, but
different functions $g_{\alpha}$ with respect to $ x $.

\begin{figure}
 \includegraphics[width=3.5in]{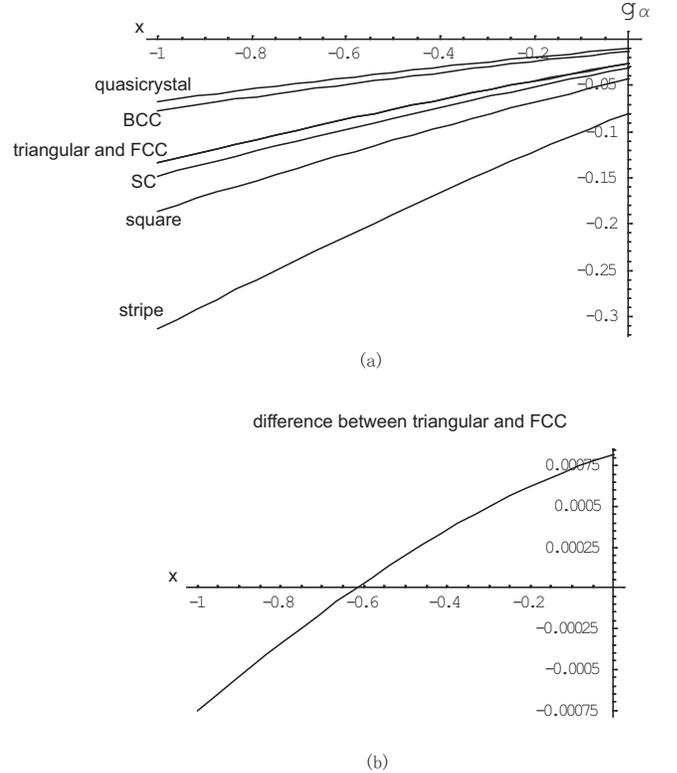}
 \caption{ $ u $ is negative. (a) $g_{\alpha}(x)$ of seven different lattices, it is hard to see the
  difference between FCC and triangular in this scale.
   (b) The difference between triangular and FCC in the expanded scale. }

\end{figure}

 Comparing $ g_{\Box},g_{\Vert},g_{\triangle},g_{bcc},g_{fcc},g_{sc},g_{qc}$ shown in Fig.8a,
 we find that there is a shift of order between triangle lattice and FCC
 lattice shwon in Fig.8(b). The transition temperature of FCC is
 $ T_{fcc}= \frac{1}{2}\frac{u_{fcc}^{2}}{v_{fcc}}=\frac{441}{1280}\frac{u^{2}}{v}$  and that of triangular lattice is
 $T_{\triangle}= \frac{1}{2}\frac{u_{\triangle}^{2}}{v_{\triangle}}=\frac{45}{136}\frac{u^{2}}{v}$. It shows that as the
 temperature is decreased, the first solid phase between these two is
 FCC, but when the temperature is further decreased below the
 transition temperature of triangular lattice and when
 $r < -0.617\frac{u^{2}}{v}$, the triangular lattice has the lower
 energy than FCC, which means that FCC is a mestable state after
 that. In general, we have the following relations,
 when $ -0.617\frac{u^{2}}{v}< r < T_{fcc}$,
 $ g_{\Vert}<g_{\Box}<g_{sc}<g_{fcc}<g_{\triangle}<g_{bcc}<g_{qc} $
 thus $f_{\Vert}<f_{\Box}<f_{sc}<f_{fcc}<f_{\triangle}<f_{bcc}<f_{qc}$.
 When $ r < -0.617\frac{u^{2}}{v} $,
 $g_{\Vert}<g_{\Box}<g_{sc}<g_{\triangle}<g_{fcc}<g_{bcc}<g_{qc}$
 thus$f_{\Vert}<f_{\Box}<f_{sc}<f_{\triangle}<f_{fcc}<f_{bcc}<f_{qc}$.
 In any case, the stripe phase is always the lowest energy
 state of all the seven lattices.

 In fact, we can get the same
 result from the critical transition temperatures of different lattices. It is known that the transition
 temperature in the above model is
 $ r_{c}=\frac{1}{2}\frac{u_{\alpha}^{2}}{v_{\alpha}}$, Plugging
 $u_{\alpha}$ and $ v_{\alpha}$ for different lattices, we find out
 that the stripe lattice has the highest transition temperature as
 expected, which means when we decrease the temperature, the first
 solid phase will be the stripe phase.

\section{Conclusions}

  In this paper, we study the transition from the normal state to the LOFF state from the GL free energy
  in a mean field theory. We consider seven most common lattices.
  By comparing the free energy and the transition temperature of the
  seven lattice structures, we find that the lowest energy lattice structure of the LOFF state is
  the stripe phase, which is the LO state originally proposed by Larkin and Ovchinnikov \cite{lo}.
  Our result shows that in heavy fermion system or cold atom system, at a sufficiently low temperature,
  if a LOEF state can be realized, then its lattice structure will likely to be a  ( stripe ) LO
  phase which will lead to anisotropy in many physical measurable
  quantities. Although so far, there is no direct probe on the structure of the order parameter
  in all these heavy fermion materials, in experiment in \cite{capan}, the thermal conductivity
measurement was used to probe the anisotropy of the order parameter,
especially the structure of the nodes in the momentum space. The
experiment indeed show the anisotropy of the thermal conductivity of
$CoCeIn_{5}$ in the possible LOFF state regime in Fig.1. Our results
suggest  that the LOFF state observed in the experiment is the
original LO state. Of course, the order parameter may contain higher
Harmonics terms. Recently, it was argued in \cite{yip} that the LO
state may be stable in an appreciable regime in  the imbalance
versus detuning phase diagram in the BCS side of the Feshback
resonance.  It is not known if the GL action still can be used to
describe the normal to the LO transition at $ T=0 $ where $ r=
p-p_{c} $ where $ p_{c} $ is the critical polarization difference,
because at $ T=0 $, the residual fermions can not be integrated out,
especially near the transition point. However, we expect the normal
to the LOFF state transition is still of the Lifshitz type first
order transition. Well inside the LOFF state, mean field analysis in
the paper still holds, so the results still apply.

   We thank Kun Yang for helpful discussions and Yong Tang for technical support. The Research at KITP
   was supported in part by the NSF under grant No. PHY-05-51164.

\appendix
 \begin{center}
    {\bf APPENDIX}
  \end{center}
\section{numerical factor in BCC and Quasicrytal lattices}

In this appendix we present in detail the procedures to get the
numerical factors for the forth and sixth order term used in the
main text. There are many ways to draw the direction of arrows in
the diagrams in the main text. Of course, all the different ways
should give exactly the same numerical factors. But for some
choices, special cares are needed to avoid overcounting the
contributions \cite{tom}. In the main text, we just showed the most
convenient choice.

 Now there are two methods to get the paired vector contribution to the forth
and sixth order term. The first method A is a constructive method by
which we count the number of ways the $ \psi( \vec{G} ) $ can take
one by one, this way is straightforward and can naturally avoid any
possible over-countings, but it is a little bit tedious especially
when the order increases. The second method is by some combination
trick method B, this way is less straightforward, but can be more
effective when the order increases. The agreement of the final
results between the two methods can insure the correctness of our
results.

{\sl Method A:}

 For the forth order term, the first $ \psi_{G} $ can take $ m $ choices and then
 (1) the second $ \psi_{G}$ takes the same vector again and then the
 next two $ \psi_{G}$ must take exact opposite of that vector, so there is only $ 1 $ choice here
 (2) the second $ \psi_{G}$ takes the opposite of the first vector.
  Then for the third and forth $ \psi_{G}$ have to be opposite and have m
  choices. This case essentially reduces to the quadratic case.
 (3) the second $ \psi_{G}$ takes one of the $ m-2 $ choices
   which is different than the first vector and its opposite.
   Then the third and forth must be the exact opposite of the first and second
   vector, therefore there are only $ 2 $ choices here.
  The total sum of all the choices are $ m[ 1+m+2(m-2) ] = 3m(m-1)$.
  After rescaling by $m^{2}$, we get $ u^{p}(m)= 3(1-\frac{1}{m})u$.

  For the sixth order term, the first $ \psi_{G} $ can take $ m $ choices and then
 (1) the second $ \psi_{G}$ takes the same vector again
  (a) the third $ \psi_{G}$  also take the same vector, then there is only
  one choice left for the rest three $ \psi_{G} $.
  (b) the third $ \psi_{G}$  also take the opposite vector, then
  from the calculations in the forth order term, then there are $ 3m-3 $
  choices for the rest three $ \psi_{G} $.
  (c) the third $ \psi_{G}$ takes one of the $ m-2 $ choices
   which is different than the first vector and its opposite,
   then there are $ 3 $ choices for the rest three $ \psi_{G} $.
   So, adding $ (a)+(b)+(c) $, there are $ 1+ (3m-3)+ 3(m-2)=6m-8 $ choices for case 1.
  (2) the second $ \psi_{G}$ takes the opposite of the first vector.
   Then this case essentially reduces to the forth order case, so
   there are $ 3m(m-1) $ choice.
  (3) the second $ \psi_{G}$ takes one of the $ m-2 $ choices
   which is different than the first vector and its opposite.
   (a) the third $ \psi_{G}$ takes one of the first two choices,
   there there are $ 2 \times 3 $ choices (b)
   the third $ \psi_{G}$ takes the opposite of one of the first two choices,
   there there are $ 2 \times 3(m-1) $ choices.
   (c) the third $ \psi_{G}$ takes one of the $ m-4 $ choices
   which is different than the first and the second vectors and their
   opposites, then there are $ 6 $ choices.
    So, adding $ (a)+(b)+(c) $, there are $ (m-2) [ 6+ 6(m-1) +6 (m-4)] =12(m-2)^{2} $ choices for case 3.
    Adding all the $ (1)+(2)+(3)= m [ 6m-8 + 3m(m-1)+ 12 (m-2)^{2}
    ]=5m ( 3m^{2}-9m+8) $. After rescaling by  $ m^{2} $, we get $
    v^{p}_{\alpha}= 5(3-9/m+8/m^{2} )v $.

{\sl Method B.}

 For the forth order term,
 There are 2 choices: (1) we choose same pair twice or choose two different pairs.
 If we choose same pair twice, first we have $\frac{m}{2}$ choices of paired vector,
 then we put this pair into 4 location. The contribution of this is $\frac{m}{2}\times\dbinom{4}{2}=3m$.
 (2) We choose two different pairs, which is $\dbinom{\frac{m}{2}}{2}$ and
 put them into 4 different location, that will be $4!$. So this term will give $3m(m-2)$. The sum of the above two contributions gives $ 3m(m-1) $.
    Rescaling by $m^{2}$, we get $ u^{p}_{\alpha}= 3 (1-1/m) u $ which is the same as that
    achieved by the method A.

 For the sixth order term: (1) we choose the same pair three times and put them into 4 locations,
 which is  $\frac{m}{2}\times\dbinom{6}{3}=10m$. (2) we have two pairs with one pair chosen twice,
 that will be $ 2\times\dbinom{\frac{m}{2}}{2}\times\dbinom{6}{2}\times\dbinom{4}{2}\times2!=45m(m-2)$.
 (3) we have three different pairs in 4 locations, which is
 $\dbinom{\frac{m}{2}}{3}\times6!=15m(m-2)(m-4)$.
 The sum of the above three contributions gives $ 5m(3m^{2}-9m+8)$.
 Rescaling by $m^{3}$, we get $ v^{p}_{\alpha}={5(3m^{2}-9m+8)}/{m^{2}}v$ which is the same as that
 achieved by the method A.

Next, we are going to show how to get the nontrivial terms for BCC
and Quasicrytal.

{\sl 1. BCC lattice}

 For BCC, we can see from Fig. 4 that for each
vertex, there are two arrows coming in and two arrows coming out
which, in spin ice case, is called "two in, two out" rule and the
sum of four vectors from
any vertex must be equal to 0. \\
{\sl a. The forth order term:} So we have this nontrivial vertex
contribution to the forth order term. There are 6 vertices, and
therefore there are 6 sets of these vectors. Their contribution to
the forth order term after rescaling
is $ u_{vet}= u $. \\
{\sl b. The sixth order term: } In addition to the paired
contributions calculated by the two methods above, there
are three non-paired contributions listed in Fig.5.\\
(5a) there is a paired vectors plus four vectors coming out of any of the 6
vertices. There are six pairs of vectors and 6 vertices.If we choose any vertice, there are 4 pairs of vectors we have exactly one vector already chosen inside of the vertice So the contribution of this $ 6\times 4 \times \dbinom{6}{2}\times 4!=8640$. And also we have 2 pairs having no same vector as that in this vertices.. This contribution is $6\times2\times6!=8640$. The sum of these two terms after rescaling gives $10v$.\\
(5b) two different triangles having a common edge. For
each edge, we have exactly one of these choices, so there are going
to be 12. We want to put one triangles into 6 locations. The number of way to do that is $ 12\times \dbinom{6}{2}\times 4!=4320$.After rescaling,we have $\frac{5}{2}v$   \\
(5c) by observation the sum of three vectors from a closed triangles equals to 0,
 therefore there is a contribution coming from one closed triangle with each side chosen twice and there
  are 8 different closed triangles. The number of ways to do that is $ 8 \times \dbinom{6}{2}\times \dbinom{4}{2}=720$. After rescaling,we have $\frac{5}{6}v$\\
  Note that two triangles having no common edge contribution has already been
  included in the (5b) and (5c).
  After the sum of (5a),(5b),(5c) and rescale, we get $ v_{bcc}=220/9v $.\\

{\sl 2. Quasicrytal lattice}

{\sl a. The forth order term:} Following \cite{tom}, in addition to
the
 paired vector contribution calculated by the methods above, there
 are also 30 non-planar diamond contribution(Fig.6(a)) to the forth order term.After rescaling, we have $u_{upd}=\frac{4}{5}u$.
 \\
{\sl b. The sixth order term:} Following the same procedure for the
BCC lattice, in addition to the paired vectors contribution, there
are four non-paired contributions listed in Fig.6. (6a) one paired
vectors plus any non-planar diamond structure. There are
15 pairs of vectors and 30 non-planar diamond structure. Their contribution after rescaling is $ 30\times4 \times\dbinom{6}{2}\times4!+30\times(15-4)\times6!=280800$. After rescaling, we have $\frac{52}{5}v$.\\
(6b) we have non-planar closed triangles with a common edge. For each edge,
there is exactly one such configuration, so there are 30 of them. Their contribution is $ 30 \times \dbinom{6}{2}\times 4!=10800$.After rescaling, we have $\frac{2}{5}v$. \\
(6c) there is closed triangle with each sides chosen twice. In
Qccrytal, there are 20 different closed triangles. Their contribution is $ 20 \times \dbinom{6}{2}\times \dbinom{4}{2}=1800$. After rescaling, we have $\frac{1}{15}v$\\
(6d) a contribution from two different triangles with no common edges. For each triangles, there are 12 different triangles that haven't been
included in previous contributions.Their contribution is $20\times12\times6!/2=86400$. After rescaling, we have $\frac{16}{5}v$\\
 After the sum of all these terms plus the trivial contribution from paired vectors, we get $ v_{qc}=497/18v $.

\section{ Liquid to solid transition, revisit }

 The liquid to solid transition was studied in \cite{tom} by
 considering only the cubic term. In this appendix, we will consider
 the effects of both the cubic and forth order term.
For liquid to solid transition, expanding the order parameter to the
forth order term, we have
\begin{eqnarray}
f_{n} & = &  \sum_{\vec{G}} \frac{1}{2}  r_{\vec{G}} | n_{\vec{G}}
|^{2} -  w  \sum_{\vec{G}} n_{\vec{G}_1}
    n_{\vec{G}_2} n_{\vec{G}_3} \delta_{ \vec{G}_1 + \vec{G}_2 +
    \vec{G}_3 ,0 }    \nonumber \\
   & + &  u  \sum_{\vec{G}} n_{\vec{G}_1}
    n_{\vec{G}_2} n_{\vec{G}_3} n_{\vec{G}_4} \delta_{ \vec{G}_1 + \vec{G}_2 +
    \vec{G}_3 + \vec{G}_4,0 }   + \cdots
\label{compare}
\end{eqnarray}
   Obviously, the difference between liquid to solid transition and the
   normal state to LOFF state transition considered in the main test is
   that there is a cubic term in the former, but not in the latter.
     Following \cite{tom}, one can simplify Eqn.\ref{compare} to:
\begin{equation}
  f_{\alpha} = \frac{1}{2} r_{\vec{G}} | n_{\vec{G}} |^{2}
  - w_{\alpha}| n_{\vec{G}}|^{3} + u_{\alpha} | n_{\vec{G}}|^{4} + \cdots \label{simple}
\end{equation}

 Because the Quantum Hall to insulator transition in single layer quantum Hall system \cite{cbtwo},
 Excitonic superfluid to Excitonic solid transition in electron-hole
 bilayer system \cite{ess} happen in two dimensions,
we will first compare two dimensional lattices, namely stripe lattice,
 square lattice and triangular lattice.  For stripe and square lattice, it is easy to see the cubic
term  $ w_{\alpha} =0 $, because there is no closed triangle in
all these lattices. Minimizing the free energy we have
$f_{\alpha}=-\frac{r^{2}}{16u_{\alpha}}$. Since $u_{\Vert}<u_{\Box}$, it is very easy to see that $f_{\Vert}<f_{\Box}$.\\
For triangular lattice, the contribution to the cubic term from a
closed triangle was evaluated in \cite{tom} to be $ w_{\triangle}=
\frac{4}{\sqrt{6}}w $. From the appendix A, we get $
u_{\triangle}=2.5u $.  Minimizing the free energy Eqn.\ref{simple}
leads to:
\begin{eqnarray}
n_{\triangle}&=&\frac{-3w_{\triangle}+\sqrt{9w_{\triangle}^{2}-16u_{\triangle}r}}{8u_{\triangle}}\nonumber\\
f_{\triangle}&=&\frac{1}{4}(-\frac{w_{\triangle}r}{4u_{\triangle}}-(r-\frac{3w_{\triangle}^{2}}{4u_{\triangle}})\frac{3w_{\triangle}}{4u_{\triangle}})n_{\triangle}-\frac{1}{4}(r-\frac{3w_{\triangle}^{2}}{4u_{\triangle}})\frac{r}{4u_{\triangle}} \nonumber \\
\end{eqnarray}
 We can see although the stripe lattice doesn't have a cubic term,
 but $ u_{\Vert} <  u_{\triangle} $. The more complete way to evaluate  which one has a lower free
energy must take these two terms into consideration, not just
considering the cubic term as did in \cite{tom}. Now Define
$r=\frac{w^{2}}{u}x$ and find out the difference between
$f_{\triangle},f_{\Vert}$, we compare numerically the two functions
within the range $ r<0 $. We still find that the triangular lattice
always has a lower free energy than stripe lattice when it is close
to the transition point as shown in Fig.9. When the temperature is
further decreased, we find that numerically $ f_{\triangle}>
f_{\Vert} $. But it is known that GL theory is only valid for weak
first order transition and second order transition close to
transition point. If the temperature is further decreased, the
validity of Eqn.\ref{simple} may be questioned. As shown in Fig.9,
we still find that the triangular lattice always has a lower free
energy than square lattice.
\begin{figure}
 \includegraphics[width=3.5in]{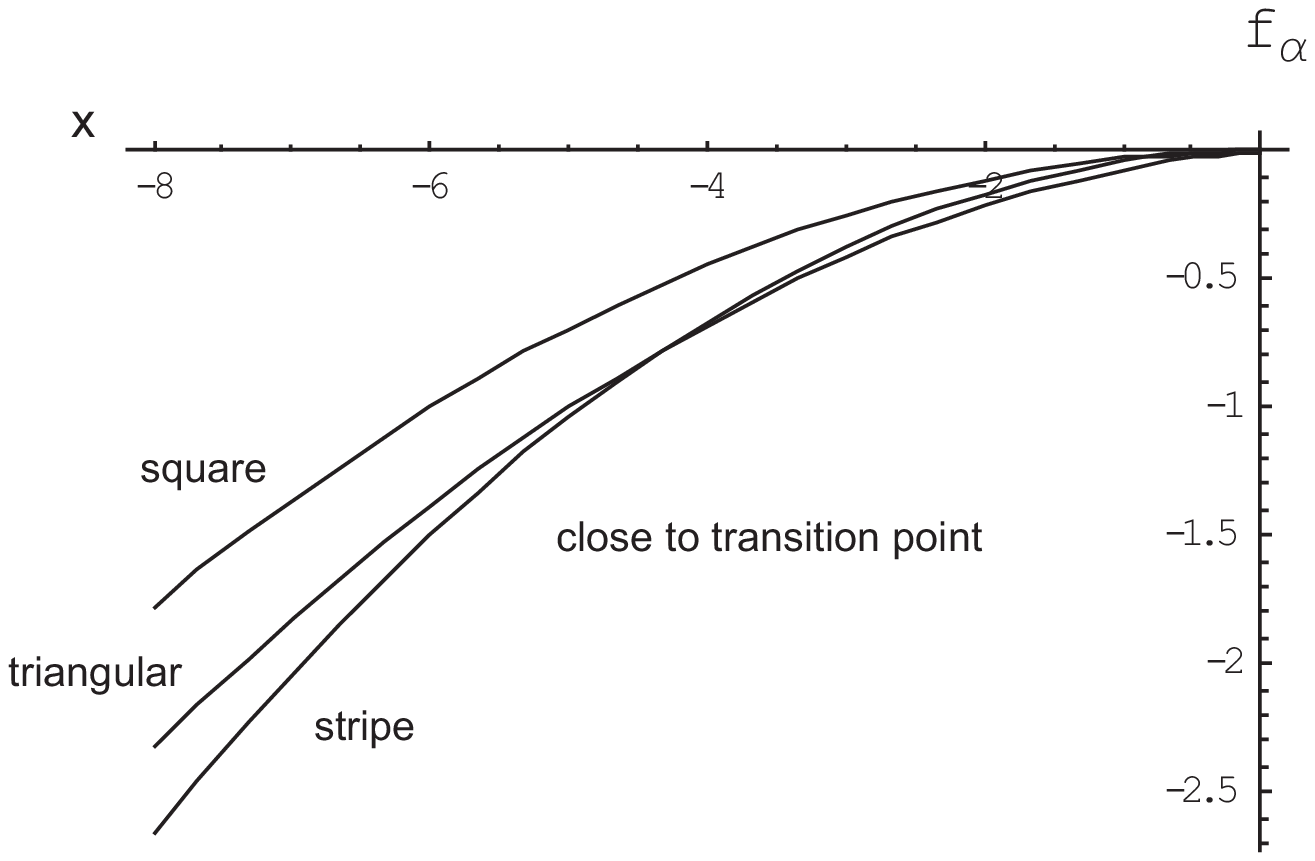}
 \caption{ Function $ f $ for 3 two dimensional lattices: $ f_{\Box}, f_{\triangle}, f_{\Vert}$ }

\end{figure}
\\

Now we will generalize the above consideration to the seven lattices
listed in table I. For stripe, square, SC and FCC, it is easy to see
the cubic term  $ w_{\alpha} =0 $, because there is no closed
triangle in all these lattices. Minimizing the free energy we have
$f_{\alpha}=-\frac{r^{2}}{16u_{\alpha}}$.

For triangular lattice, BCC and Quasicrystal, the contribution to
the cubic term from a closed triangle was evaluated in \cite{tom} to
be $ w_{\alpha}=4/ \sqrt{m}$ after rescaling. Minimizing the free
energy Eqn.\ref{simple} leads to:
\begin{eqnarray}
n_{\alpha}&=&\frac{-3w_{\alpha}+\sqrt{9w_{\alpha}^{2}-16u_{\alpha}r}}{8u_{\alpha}}\nonumber\\
f_{\alpha}&=&\frac{1}{4}(-\frac{w_{\alpha}r}{4u_{\alpha}}-(r-\frac{3w_{\alpha}^{2}}{4u_{\alpha}})\frac{3w_{\alpha}}{4u_{\alpha}})n_{\alpha}-\frac{1}{4}(r-\frac{3w_{\alpha}^{2}}{4u_{\alpha}})\frac{r}{4u_{\alpha}} \nonumber \\
\end{eqnarray}
Following the same method used previously, we can define
$r=\frac{w^{2}}{u}x$ and find the difference between $f_{\alpha}$.
We compare numerically the functions within the range $ r<0 $. Since
we know that $ u_{sc}< u_{fcc}$, $ f_{sc} < f_{fcc}$, so in order to
find  which one has the lowest energy, we only need to compare
triangular lattice, SC and Quasicrystal with BCC. In Fig.10, we only
show the difference between BCC, Quasicrystal, triangular and SC.
\begin{figure}
 \includegraphics[width=3.5in]{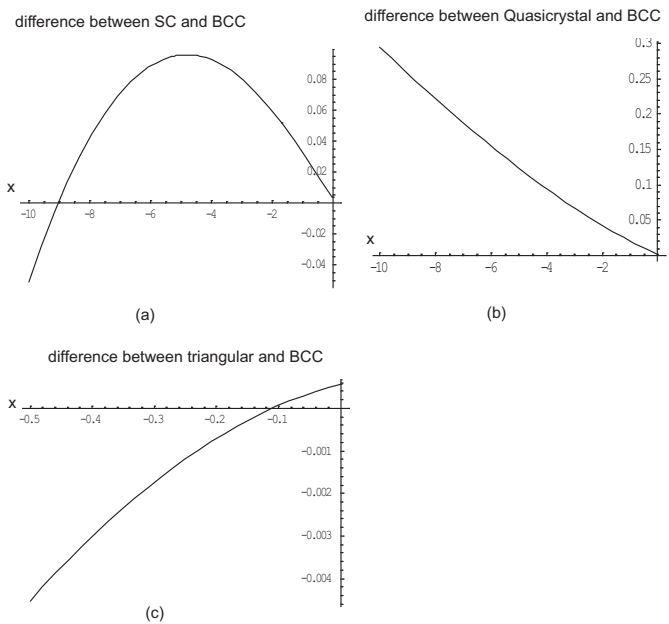}
 \caption{ Difference between (a)  $f_{sc}$ and $f_{bcc}$,
 (b)  $f_{qc}$ and $f_{bcc}$
 (c)  $f_{\triangle}$ and $f_{bcc}$ }
\end{figure}

 We find that when the temperature is decreased just below the
 transition temperature of the lattices, the lattices with a cubic
 term have a smaller free energy than the lattices which do not.
 Fig.10 shows that BCC lattice has the lowest free energy and
 the highest transition temperature in a range just below the transition point.

\end{document}